\documentclass[conference]{IEEEtran}

\usepackage{graphicx}
\usepackage{amsmath,amssymb,amsfonts}
\usepackage{multirow}
\usepackage{psfrag}
\usepackage{subfigure}
\usepackage{color}
\usepackage{soul}
\usepackage{xcolor}
\usepackage[bookmarks=false]{hyperref}
\usepackage{lipsum}
\usepackage[ruled,vlined,linesnumbered]{algorithm2e}
\usepackage{mathrsfs}
\usepackage{pbox}
\usepackage{amsthm}
\usepackage{wrapfig}

\DontPrintSemicolon

\makeatletter
\def\hlinewd#1{%
  \noalign{\ifnum0=`}\fi\hrule \@height #1 \futurelet
   \reserved@a\@xhline}
\makeatother

\ifCLASSINFOpdf
\else
\fi

\DeclareMathOperator*{\minimize}{minimize}

\begin{document}

\title{UBAT: On Jointly Optimizing UAV Trajectories and Placement of Battery Swap Stations}

\author{Myounggyu Won\\
Department of Computer Science, University of Memphis, Memphis, TN, United States\\
mwon@memphis.edu}%


\markboth{Journal of \LaTeX\ Class Files,~Vol.~13, No.~9, September~2014}%
{Shell \MakeLowercase{\textit{et al.}}: Bare Demo of IEEEtran.cls for Journals}

\maketitle

\begin{abstract}
Unmanned aerial vehicles (UAVs) have been widely used in many applications. The limited flight time of UAVs, however, still remains as a major challenge. Although numerous approaches have been developed to recharge the battery of UAVs effectively, little is known about optimal methodologies to deploy charging stations. In this paper, we address the charging station deployment problem with an aim to find the optimal number and locations of charging stations such that the system performance is maximized. We show that the problem is NP-Hard and propose UBAT, a heuristic framework based on the ant colony optimization (ACO) to solve the problem. Additionally, a suite of algorithms are designed to enhance the execution time and the quality of the solutions for UBAT. Through extensive simulations, we demonstrate that UBAT effectively performs multi-objective optimization of generation of UAV trajectories and placement of charging stations that are within 8.3\% and 7.3\% of the true optimal solutions, respectively.
\end{abstract}


\IEEEpeerreviewmaketitle

\section{Introduction}
\label{sec:introduction}

With recent breakthroughs in design and production of unmanned aerial vehicles (UAVs), UAVs are increasingly used in many applications such as military surveillance~\cite{orfanus2016self}, disaster response~\cite{erdelj2017help}, oil gas pipe inspection~\cite{hausamann2005monitoring}, precision agriculture~\cite{tokekar2016sensor}, and delivery of goods~\cite{murray2015flying}. The Federal Aviation Administration (FAA) estimates that commercial UAVs will grow 10-fold between 2016 and 2021, and the global market revenue of UAVs will rise to \$11.2 billion by 2020~\cite{uav_trend}. One of the major obstacles that constrains the huge potential of UAVs is the limited flight time. This is a significant problem as more and more UAV applications require coverage of a large geographical area, lengthening the mission duration well beyond the battery capacity of UAVs~\cite{gong2018flight}.

Numerous solutions have been developed to extend the flight time of UAVs focusing on utilization of charging stations that recharge/replace the battery of UAVs~\cite{swieringa2010autonomous}\cite{suzuki2012automatic}\cite{lu2018wireless}. While these solutions have successfully increased the operation time of UAVs, there is still a significant knowledge gap on how to deploy charging stations to maximize the system performance.  Specifically, existing approaches are based on a simplifying assumption for deploying charging stations. Wei and Isler propose a solution based on a single charging station~\cite{wei2018coverage}. Some solutions deploy charging stations only at fixed locations~\cite{tseng2017autonomous} such as at the center of a field~\cite{trotta2018joint} and the base station~\cite{scherer2017short}. Yu \emph{et al.} develop a solution based on unmanned ground vehicle (UGV)-based charging, but these UGVs are allowed to visit only the sites to be visited by the UAVs~\cite{yu2018algorithms}. It remains as a challenge to develop a solution that finds the optimal UAV trajectories with no constraints on the number and locations of charging stations.

In this paper, we address the problem of deploying charging stations that maximizes the system performance specifically concentrating on generating optimal UAV trajectories, while minimizing the number of deployed charging stations. More specifically, we formulate the problem of jointly optimizing UAV \underline{T}rajectories and \underline{L}ocations of \underline{B}attery swap \underline{S}tations (\textbf{TLBS}) as an optimization problem, and develop an optimization framework to solve the problem. Especially, since finding a shortest trajectory for a single UAV visiting each region of interest (ROI) and returning to the base station, without even considering charging stations, is essentially the travelling salesperson problem (TSP) which is NP-Hard,  \emph{\textbf{we propose a heuristic solution called UBAT based on Ant Colony Optimization (ACO) motivated by the fact that ACO is well known to solve TSP very effectively~\cite{dorigo1997ant2}.}} The technical contributions of this paper are that the standard ACO is adapted to account for unique requirements for solving the \textbf{TLBS} problem especially in calculating the probabilities for path selection and updating the pheromone trails: (1) constrained energy of UAVs, (2) simultaneous coverage of ROIs by multiple UAVs, and (3) joint optimization of UAV trajectories and number of charging stations.

The proposed framework is fine-tuned with novel algorithms that are designed to enhance the convergence speed and the quality of solutions. More specifically, an algorithm is developed to minimize the number of candidate locations for deploying charging stations, thereby reducing the search space and lowering the computational overhead. We also design an algorithm that effectively tunes the parameters for the proposed framework to enhance the quality of solutions. Additionally, the 2-OPT local search~\cite{voudouris1999guided} is effectively incorporated to further improve the quality of solutions. Extensive simulations are conducted to evaluate the performance of UBAT as well as the supporting algorithms. The results demonstrate that the proposed approach effectively performs joint optimization of UAV trajectories and placement of charging stations that are within 8.3\% and 7.3\% of the true optimal solutions, respectively.

We define the \textbf{TLBS} problem in Section~\ref{sec:sys_model} and present the details of UBAT in Section~\ref{sec:approach} followed by descriptions of supporting algorithms in Section~\ref{sec:enhanced_aco}. Simulation results are presented in Section~\ref{sec:simulation_results}. We then conclude in Section~\ref{sec:conclusion}.



\section{TLBS Problem}
\label{sec:sys_model}

\subsection{System Model}
\label{sec:target_area}

Consider a target area $A$ with an arbitrary shape in $\mathbb{R}^2$. The target area is divided into a two-dimensional grid of square cells, \emph{i.e.,} the target area is represented as a set $A = \{a_1, a_2, ..., a_{NS}\}$ where each element is a subarea (cell), and $NS$ is the number of subareas. Among the cells are regions of interests (ROIs) denoted by a set $R=\{r_1, r_2, ..., r_{NR}\} \subseteq A$, where $NR$ is the number of ROIs. We introduce the function $\mathcal{F}_{dist} : A, A \rightarrow \mathbb{R}$ which defines the Euclidean distance between two cells. More specifically, $\mathcal{F}_{dist}(a_i, a_j)$ means the Euclidean distance between the centers of two cells $a_i$ and $a_j$.

\begin{wrapfigure}{r}{0.6\columnwidth}
\centering
\includegraphics[width=.6\columnwidth]{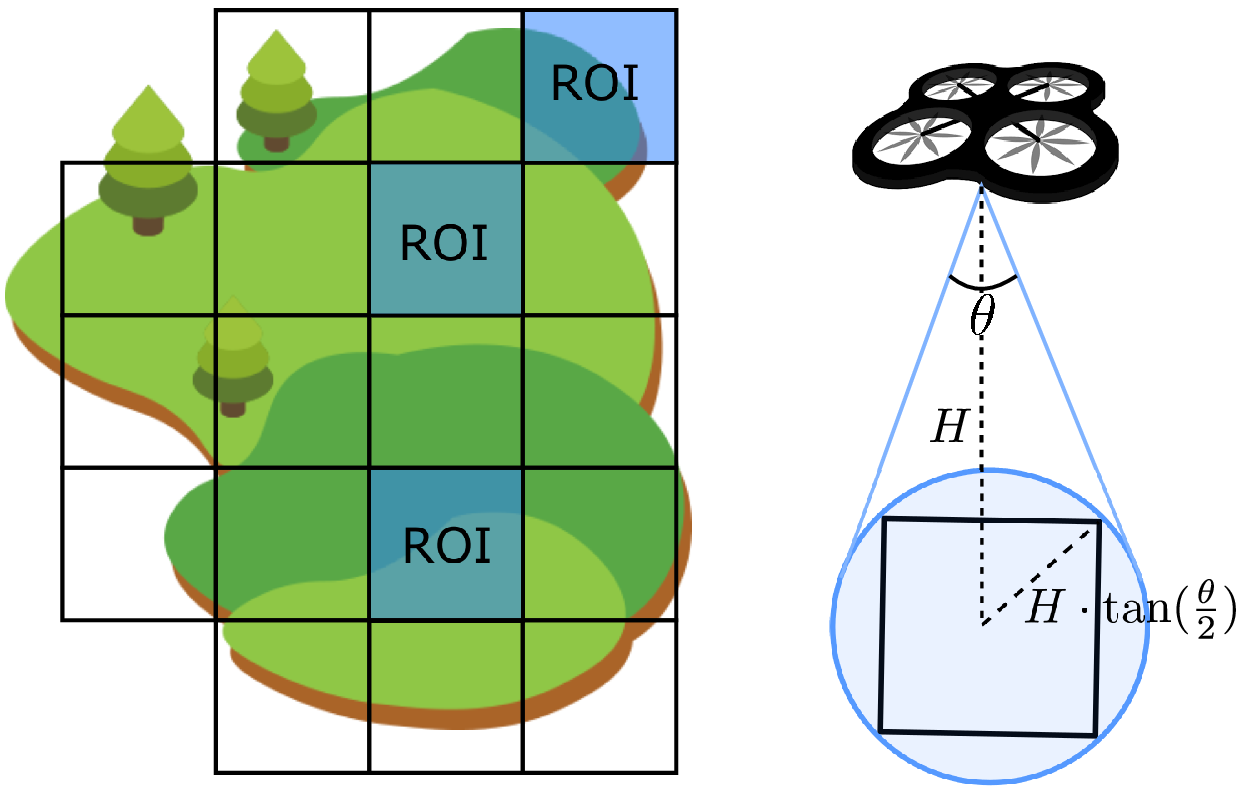}
\caption {An illustration of a target area, ROIs, and sensing coverage.}
\label{fig:target_area}
\end{wrapfigure}


There are $NU$ UAVs denoted by a set $U = \{u_1, u_2, ..., u_{NU}\}$. Assume a typical UAV application where UAVs fly over a target area, perform sensing, and transmit sensed data to the base station. We assume that UAVs fly $H$ meters high from the ground.  As such, the sensing coverage $COV$ is $\pi \cdot (H \cdot \tan (\frac{\theta}{2}))^2$ where $\theta$ is the angle of the sensing cone. The sensing coverage is sufficiently large to cover a whole ROI when a UAV is above the center of the ROI. Figure~\ref{fig:target_area} illustrates a target area divided into cells, three ROIs in the target area, and the sensing coverage. The time is discretized into time slots $T=\{t_1, t_2, ...\}$ where the length of a time slot is denoted by $LT$. In particular, let $t_{finish}$ be the time slot when all ROIs have been fully covered.

The flight time of a UAV is limited due to its energy capacity. To support continuous operation of UAVs, $NC$ charging stations are deployed in the target area which are denoted by a set $C = \{c_1, c_2, ..., c_{NC}\}$. More precisely, we assume that these charging stations automatically replace the battery of a UAV when it is landed on the charging station.



Let $E_{MAX}$ be the maximum energy capacity of a UAV. Also, let $E_{u_i, t_j}$ be the remaining energy of a UAV $u_i \in U$ at time slot $t_j \in T$. Assume that all UAVs are fully charged before operation, \emph{i.e.,} $E_{u_i, t_0} = E_{MAX}, \forall u_i \in U$. We define $\gamma_{fly}$ be the energy loss rate while a UAV is flying, which can be updated based on the factors that affect energy consumption such as wind. Each UAV $u_i$ is in one of these states $S=\{s_{fly}, s_{rec}\}$ at each time slot. The state $s_{fly}$ indicates that a UAV is flying, and the state $s_{rec}$ indicates that a UAV is landed on a charging station to get its battery replaced.

A function $\mathcal{F}_{fly} : U, A, T \rightarrow \{0, 1\}$ is defined to specify whether a UAV is flying over a certain cell at a certain time slot. For example, $\mathcal{F}_{fly}(u_i, a_j, t_k) = 1$ if UAV $u_i$ is flying over cell $a_j$ at time slot $t_k$, and $\mathcal{F}_{fly}(u_i, a_j, t_k) = 0$ if it is landed on a charging station located in a cell $a_j$. The residual energy of a UAV $u_i$ at time slot $t_j$, \emph{i.e.,} $E_{u_i, t_j}$ can then be defined as follows.

\begin{equation}
\label{eq1}
\begin{aligned}
E_{u_i, t_k} &= E_{u_i, t_{k-1}} - \gamma_{fly} \cdot \mathcal{F}_{fly}(u_i, a_j, t_k) \\
& \quad + (E_{MAX} - E_{u_i, t_{k-1}}) \cdot (1 - \mathcal{F}_{fly}(u_i, a_j, t_k)).
\end{aligned}
\end{equation}

\noindent Here, $E_{u_i, t_{j-1}}$ is the residual energy at the previous time slot. $\gamma_{fly} \cdot f_{fly}(u_i, t_j)$ represents the energy loss for flying during the time slot, and $ (E_{MAX} - E_{u_i, t_{j-1}}) \cdot (1 - f_{fly}(u_i, t_j))$ indicates replacement of the battery during the time slot.

We introduce the function $\mathcal{F}_{uloc} : U, A, T \rightarrow \{0, 1\}$ indicating whether the UAV is located in a certain cell at a certain time slot. For example, $\mathcal{F}_{uloc}(u_i, a_j, t_k) = 1$ means that UAV $u_i \in A$ is in a cell $a_j \in A$ at time slot $t_k \in T$, and otherwise $\mathcal{F}_{uloc}(u_i, a_j, t_k) = 0$. We also define the function $\mathcal{F}_{cloc} : A \rightarrow \{0, 1\}$ indicating that a charging station is deployed in a certain cell. More precisely, $\mathcal{F}_{cloc}(a_i) = 1$ if a charging station is in the cell $a_i \in A$, otherwise $\mathcal{F}_{cloc}(a_i) = 0$. Let us define another function $\mathcal{F}_{mov} : U, T \rightarrow \mathbb{R}$ which indicates the Euclidian distance moved during a time slot. For example $\mathcal{F}_{mov}(u_i, t_j)$ means the distance moved by UAV $u_i \in U$ during time slot $t_j \in T$. Let $v_{MAX}$ be the maximum speed of the UAV.

\subsection{Problem Formulation}
\label{sec:problem}

Having defined all notations and functions, the problem of jointly optimizing UAV \underline{T}rajectories and \underline{L}ocations of \underline{B}attery swap \underline{S}tations (\textbf{TLBS}) is formulated as follows.

\begin{equation}
\label{objective}
\minimize \quad \{t_{finish}, \mbox{ } \sum\limits_{i=1}^{NS}\mathcal{F}_{cloc}(a_i) \} \\
\end{equation}


\begin{equation}
\label{constraint2}
\sum\limits_{j=1}^{NS} \mathcal{F}_{uloc}(u_i, a_j, t_k) \le 1, \forall u_i \in U, \forall t_k \in T
\end{equation}

\begin{equation}
\label{constraint2.1}
\sum\limits_{i=1}^{NU}\sum\limits_{k=1}^{finish} \mathcal{F}_{uloc}(u_i, a_j, t_k) \ge 1, \forall a_j \in R
\end{equation}

\begin{equation}
\label{constraint3}
0 < E_{u_i, t_j} \le E_{MAX}, \forall u_i \in U, \forall t_j \in T
\end{equation}



\begin{equation}
\label{constraint6}
\mathcal{F}_{mov}(u_i, t_j) \le LT \cdot v_{MAX}, \forall u_i \in U, \forall t_j \in T
\end{equation}

\begin{equation}
\begin{aligned}
\label{constraint7}
\mathcal{F}_{cloc}(a_j) + \mathcal{F}_{fly}(u_i, a_j, t_k) \ge 1,\\
\forall u_i \in U, \forall a_j \in A, \forall t_k \in T
\end{aligned}
\end{equation}

\begin{equation}
\label{constraint8}
\mbox{Equation \ref{eq1}}
\end{equation}


Constraint~\ref{constraint2} asserts that each UAV can be located in only one cell at each time slot. Constraint~\ref{constraint2.1} means that each ROI is covered by at least one UAV when the UAV operation is finished. Constraint~\ref{constraint3} states that each UAV does not deplete its energy completely during operation, and the residual energy should not be greater than the maximum energy capacity. Constraint~\ref{constraint6} assures that each UAV does not move faster than the maximum speed $v_{MAX}$. Constraint~\ref{constraint7} means that if a UAV is not flying, \emph{i.e.,} landed in a cell for recharging, there should be a charging station in that cell. Finally, Constraint~\ref{constraint8} specifies how the residual energy of each UAV is updated at each time slot. The \textbf{TLBS} problem is NP-Hard because it contains, as a simplified instance, the traveling salesman problem (TSP)~\cite{dorigo1997ant}. As such, we develop a heuristic framework based on the ant colony optimization (ACO) motivated by the fact that ACO is particularly effective in solving the TSP problem~\cite{dorigo1997ant2}.

\section{Proposed Approach}
\label{sec:approach}


\subsection{Path Selection}
\label{sec:path_selection}

We represent UAVs and ROIs as ants and food sources, respectively. For simplicity, we assume that UAVs fly at a speed of $v_{MAX}$ (Constraint~\ref{constraint6}), which can be easily replaced by a vector to represent the time-varying speed. We follow the energy model defined in Eq.~\ref{eq1} to respect Constraint~\ref{constraint8}. Compared to ACO with ants, we leverage two advantages of UAVs. First, UAVs have memory; they can keep track of visited ROIs so that they will visit only ROIs that have not been covered by any other UAV. Second, UAVs have computational capabilities; they can compute the distance between two ROIs. Based on these advantages, the probability that a UAV in cell a $a_i$ selects a ROI $a_j$ to visit is defined as follows.

\begin{equation}
\label{eq:path_selection}
p_{ij} =
  \begin{cases}
    \frac{\tau_{ij}^{\alpha} \cdot \eta_{ij}^{\beta}}{\sum_{k \in R_{visit}} \tau_{ik}^{\alpha} \cdot \eta_{ik}^{\beta}}       & \quad \text{if } a_j \in R_{visit}\\
    0  & \quad \text{otherwise}
  \end{cases}
\end{equation}

\noindent Here $R_{visit}$ is the set of ROIs that have not been visited by any UAV (Constraint~\ref{constraint2.1}). $\tau_{ij}$ is the pheromone intensity of an edge connecting two cells $a_i$ and $a_j$. $\eta_{ij}$ is the inverse of the Euclidean distance between two cells $a_i$ and $a_j$ namely $\frac{1}{\mathcal{F}_{dist}(a_i, a_j)}$, which is used to ensure that a shorter edge is selected in finding a path. Two parameters $\alpha$ and $\beta$ are introduced to adjust the influence of the pheromone and the distance between two cells, respectively. After calculating the probabilities for all ROIs, a UAV selects a ROI with the highest probability and moves to the selected ROI.

If a UAV does not have enough energy to reach the selected ROI, the state of the UAV is changed to the charging mode, and the UAV finds a set of cells that it can reach with its residual energy (Constraint~\ref{constraint3}). A cell is selected from those candidate cells using a slightly modified probability model compared to Eq.~\ref{eq:path_selection}, and then the UAV moves to the selected cell in which a charging station is placed (Constraint \ref{constraint7}). Specifically for the modified probability model, $\eta_{ij}$ is set to the Euclidean distance between two ROIs (i.e., $\eta_{ij} = \mathcal{F}_{dist}(a_i, a_j)$), instead of the inverse of it as specified in Eq.~\ref{eq:path_selection}, to minimize the number of deployed charging stations by assuring that a longer edge is preferred in the charging mode, \emph{i.e.,} we allow UAVs to fly as far as possible with the given residual energy because the battery will be replaced anyways.

\subsection{Trail Update}
\label{sec:trail_update}

The pheromone trails are updated in two ways: evaporation and reinforcement. The former mimics the natural evaporation of pheromone. More specifically, pheromone $\tau_{ij}$ between two cells $a_i$ and $a_j$ are reduced if $\tau_{ij} > 0$ based on the formula: $\tau_{ij} = (1 - \rho) \cdot \tau_{ij} + \rho \cdot \tau_0$, where $\rho$ is a parameter that controls the speed of pheromone reduction, and $\tau_0$ is the initial pheromone. Evaporation is particularly useful for preventing one path from dominating other possible solutions. Due to evaporation, other paths are explored, potentially leading to a better solution.


Once a path $P$ is discovered, pheromone trails on that path are strengthened, which is called the reinforcement. The reinforcement process is used to encourage the use of shorter paths and to increase the chance of using the edges of the currently known shortest path in finding potentially better paths subsequently. Formally, the pheromone trail of each edge of the selected path is reinforced as follows: $\tau_{ij} = (1 - \rho) \cdot \tau_{ij} + \rho \cdot \frac{Q}{L}$, where $L$ is the length of path $P$; $Q$ is a system parameter -- Details on determining an appropriate value for the parameter is discussed in Section~\ref{sec:effect_of_enhancement}.


\subsection{Multiple UAVs with Minimum Number of Charging Stations}
\label{sec:multiple_uavs}

We now extend the algorithm to account for general scenarios where multiple UAVs collaborate to cover ROIs with an objective of minimizing the number of deployed charging stations. More specifically, each UAV $u_i$ visits a portion of ROIs and ends up with having an individual path denoted by $P_{u_i}$. Denote the length of path $P_{u_i}$ for UAV $u_i \in U$ be $L_{u_i}$. Then, to minimize $t_{finish}$, we try to minimize the maximum path length of all UAVs, \emph{i.e.,} $\max\limits_{1 \le i \le NU}L_{u_i}$, rather than minimizing $L$. Consequently, the pheromone updating formula is rewritten as follows: $\tau_{ij} = (1 - \rho) \cdot \tau_{ij} + \rho \cdot \frac{Q}{\max\limits_{1 \le i \le NU}L_{u_i}}$.



We then take into account the number of deployed charging stations in the pheromone updating rule for reinforcement. If we do not constrain the number of charging stations in finding a path, it may increase arbitrarily. To this end, UBAT tries to make as many of the charging stations be shared by multiple UAVs such that the path length is also optimized. The idea to implement this is to include the number of deployed charging stations $NC$ in the pheromone updating formula. More specifically, we aim to find cells to deploy charging stations such that both $t_{finish}$ and the number of deployed charging stations $NC$ are minimized, as specified in the objective function (Eq.~\ref{objective}). In order to encourage the use of a path that minimizes both $t_{finish}$ and $NC$, we can rewrite the pheromone updating formula as follows: $\tau_{ij} = (1 - \rho) \cdot \tau_{ij} + \rho \cdot (\frac{Q_1}{\max\limits_{1 \le i \le NU}L_{u_i}} + \frac{Q_2}{NC})$.



\begin{algorithm}
  \label{algorithm1}
  \caption{ACO for TLBS problem}
  \Begin {
            Initialize $\alpha$, $\beta$, $\tau_0$, $Q_1$, $Q_2$ and $\rho$.\;
            \While{time $<$ MAX\_ITERATION}{
                      \While{$R_{visit} \neq \emptyset$}{
                                \For{each $u \in U$}{
                                          Find a cell $r \in R$ to visit using Eq.~\ref{eq:path_selection}.\;
                                          \uIf{$r$ is reachable with $E_u - E_{threshold}$}{
                                                    $R_{visit} \leftarrow R_{visit} - \{r\}$.\;
                                                    $c \leftarrow $ current cell; $P_{u} \leftarrow P_{u} \cup Seg_{cr}$.\;
                                          }
                                          \uElse{
                                                   Find a set of cells $\hat{A} \subseteq A$ that are reachable with residual energy.\;
                                                   Locate a cell $a \in \hat{A}$ to place a charging station using the modified version of Eq.~\ref{eq:path_selection}.\;
                                                   $c \leftarrow $ current cell; $P_{u} \leftarrow P_{u} \cup Seg_{ca}$.\;
                                                   $NC \leftarrow NC + 1$.\;
                                          }
                                }
                      }
                      Update $\tau_{ij}$ based on evaporation $\forall i, j \in A$ if $Seg_{ij} \notin \bigcup\limits_{u_i \in U} P_{u_i}$.\;
                      Update $\tau_{ij}$ based on reinforcement $\forall i, j \in A$ if $Seg_{ij} \in \bigcup\limits_{u_i \in U} P_{u_i}$.\;
                      $P_{u} \leftarrow \emptyset$, $\forall u \in U$.\;
            }
  }
\end{algorithm}

\setlength{\textfloatsep}{1pt}

Algorithm~\ref{algorithm1} summarizes the operation of UBAT. Starting from the current cell $c \in A$, each UAV $u \in U$ selects the next cell $r \in R$ to visit using Eq.~\ref{eq:path_selection} (Line 6). If the selected cell is reachable with the remaining energy $E_{u}$, the edge connecting $c$ and $r$, namely $Seg_{cr}$ is added to the path $P_u$ of UAV $u$ (Lines 7-9). On the other hand, if $r$ is not reachable, the algorithm locates a cell $a$ to place a charging station based on the modified version of the probability model Eq.~\ref{eq:path_selection} (Lines 11-14). More precisely, a charging station is selected such that both $NC$ and $\max\limits_{1 \le i \le NU}L_{u_i}$ are minimized, and the path segment connecting cells $c$ and $a$ becomes part of $P_u$. Since a new charging station has been placed in cell $a$, the number of charging stations is incremented by one. This process is repeated until all ROIs have been visited by UAVs, and as a result, we obtain the trajectories for all UAVs and the cells that have charging stations. Given the discovered paths $P_u, \forall u \in U$, the pheromone trails are updated based on evaporation and reinforcement (Lines 15-16). This whole process is repeated until we reach $MAX\_ITERATION$ (Line 3).

\section{Enhancement Techniques}
\label{sec:enhanced_aco}



\textbf{Reducing Search Space:} We develop an algorithm to improve the execution time of UBAT by reducing the overhead for finding a set of candidate cells $\hat{A}$ for deploying a charging station (Lines 10-14 of Algorithm~\ref{algorithm1}). The key idea is to make the search space smaller based on the observation that a charging station can be deployed only within the convex hull of ROIs. Consequently, only the cells within the convex hull are considered in finding $\hat{A}$.


\begin{figure}[!htbp]
\centering
\begin{minipage}[b]{0.405\columnwidth}
\centering
\includegraphics[width=\columnwidth]{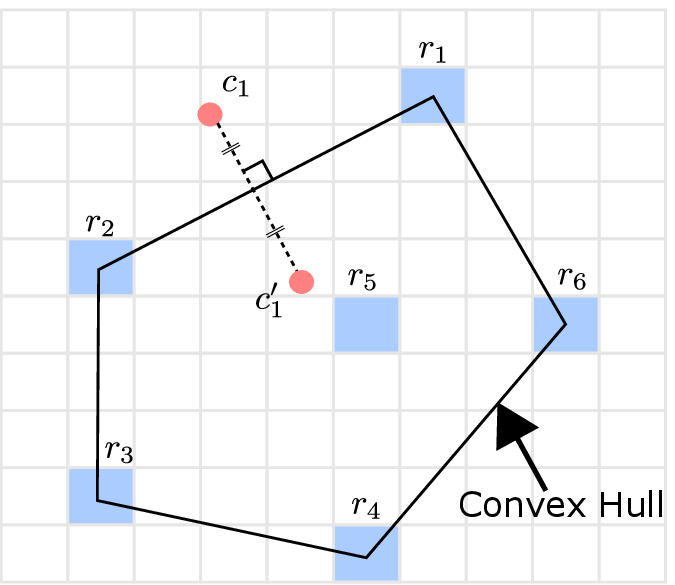}
\caption {An illustration of candidate cell selection.}
\label{fig:cs_selection_ex}
\end{minipage}
\hspace{8mm}
\begin{minipage}[b]{0.405\columnwidth}
\centering
\includegraphics[width=\columnwidth]{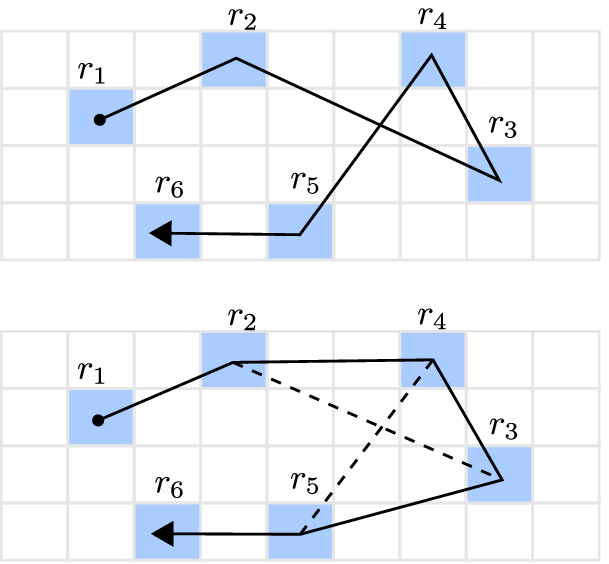}
\caption {An illustration of the 2-OPT local search.}
\label{fig:2opt}
\end{minipage}
\end{figure}


We can easily prove that deploying charging stations only inside the convex hull of ROIs does not influence the quality of the solution. Assume in contradiction that a charging station is placed outside the convex hull, \emph{e.g.,} $c_1$ in Figure~\ref{fig:cs_selection_ex} to allow the UAV to move from $r_1$ to $r_2$. We can always find $c_1'$ such that $\mathcal{F}_{dist}(r_1, c_1) + \mathcal{F}_{dist}(c_1, r_2) = \mathcal{F}_{dist}(r_1, c_1') + \mathcal{F}_{dist}(c_1', r_2)$. Furthermore, $c_1'$ is a better choice because it is closer to other ROIs considering that it may be used by other UAVs.


\textbf{Tuning Parameters:} Another key factor that affects the performance of UBAT is the parameters $Q_1$ and $Q_2$. If we choose too large values for $Q_1$ and $Q_2$, \emph{i.e.,} $\tau_0 << \frac{Q_1}{\max\limits_{1 \le i \le NU}L_{u_i}} + \frac{Q_2}{NC}$, the pheromone intensity of a new path would become too strong. As a result, it will take long for the pheromone of a newly found path to evaporate down to the level that other potential paths can be explored. On the other hand, if the values for the parameters are too small, \emph{i.e.,} $\tau_0 >> \frac{Q_1}{\max\limits_{1 \le i \le NU}L_{u_i}} + \frac{Q_2}{NC}$, the pheromone intensity of a new path would become too small leading to the situation where a potentially good solution is dropped too early, and as such, UBAT misses a chance to validate that it was actually a good solution.

We aim to find adequate values for $Q_1$ and $Q_2$ such that $\tau_0 \approx \frac{Q_1}{\max\limits_{1 \le i \le NU}L_{u_i}} + \frac{Q_2}{NC}$. To this end, we adopt an iterative approach. More specifically, we set $\tau_0$ to 1 and start UBAT with arbitrary values for $Q_1$ and $Q_2$. After only a few iterations, we get intermediate solutions $a$ and $b$ for $\max\limits_{1 \le i \le NU}L_{u_i}$ and $NC$, respectively. We use these intermediate solutions as new values for the parameters (\emph{i.e.,} $Q_1 \leftarrow a, Q_2 \leftarrow b$), and then, we repeat the same process to obtain new values for the parameters. Through experiments in Section~\ref{sec:effect_of_enhancement}, we validate that this approach is quick and effective requiring only one or two iterations for finding adequate values of the parameters.


\textbf{Improving paths:} To make UBAT run faster and obtain a higher quality solution, we perform correction on a path found in each iteration by applying the 2-OPT local search algorithm~\cite{voudouris1999guided}. The algorithm was first proposed by Croes for improving the quality of a solution for TSP~\cite{croes1958method}. The idea is to detect a path that crosses over itself and perform reordering of cells so that the cross over is removed. More precisely, two edges that cross each other are removed from a path, and then the resulting two path segments after removal of the edges are reconnected such that the cross over no longer exists. For example in Figure~\ref{fig:2opt}, an edge $r_2 \rightarrow r_3$ is replaced by $r_2 \rightarrow r_4$; and an edge $r_4 \rightarrow r_5$ is replaced by $r_3 \rightarrow r_5$, resulting in a shorter path with no cross over. Unlike the traditional 2-OPT local search, however, we ensure that the output of the algorithm does not result in increased path length for other UAVs and deployment of additional charging stations.

\section{Evaluation}
\label{sec:simulation_results}

We implemented UBAT using C++ on a PC equipped with a 2.5GHz dual-core Intel processor, and 16GByte RAM. A field with dimensions of 20$\times$20km$^2$ was created which was divided into cells with dimensions of 1$\times$1km$^2$, \emph{i.e.,} 400 cells in the field. Random scenarios were created by selecting 10 ROIs randomly. UAVs started operation from one of the randomly selected ROIs. We conducted simulations with different numbers of UAVs. The batteries of all UAVs were fully charged. The maximum distance that a fully charged UAV can fly was set to 5km. The sensing range of a UAV was configured to fully cover a cell when it hovers over the cell. The parameters for ACO were set as $\alpha = 2, \beta = 2$, and $\rho = 0.3$. The max iteration count was set to 30,000. The main metrics measured were (1) the length (km) of the longest path of all UAVs, \emph{i.e.,} $\max\limits_{1 \le i \le NU}L_{u_i}$ (which will be referred to as `path length' hereafter), and (2) the number of deployed charging stations (CS). The path length and number of CS of UBAT were compared with the true optimal solutions. The effectiveness of the proposed enhancement techniques was also evaluated.

\begin{figure}[!htbp]
\begin{minipage}[b]{0.485\columnwidth}
\centering
  \includegraphics[width=\linewidth]{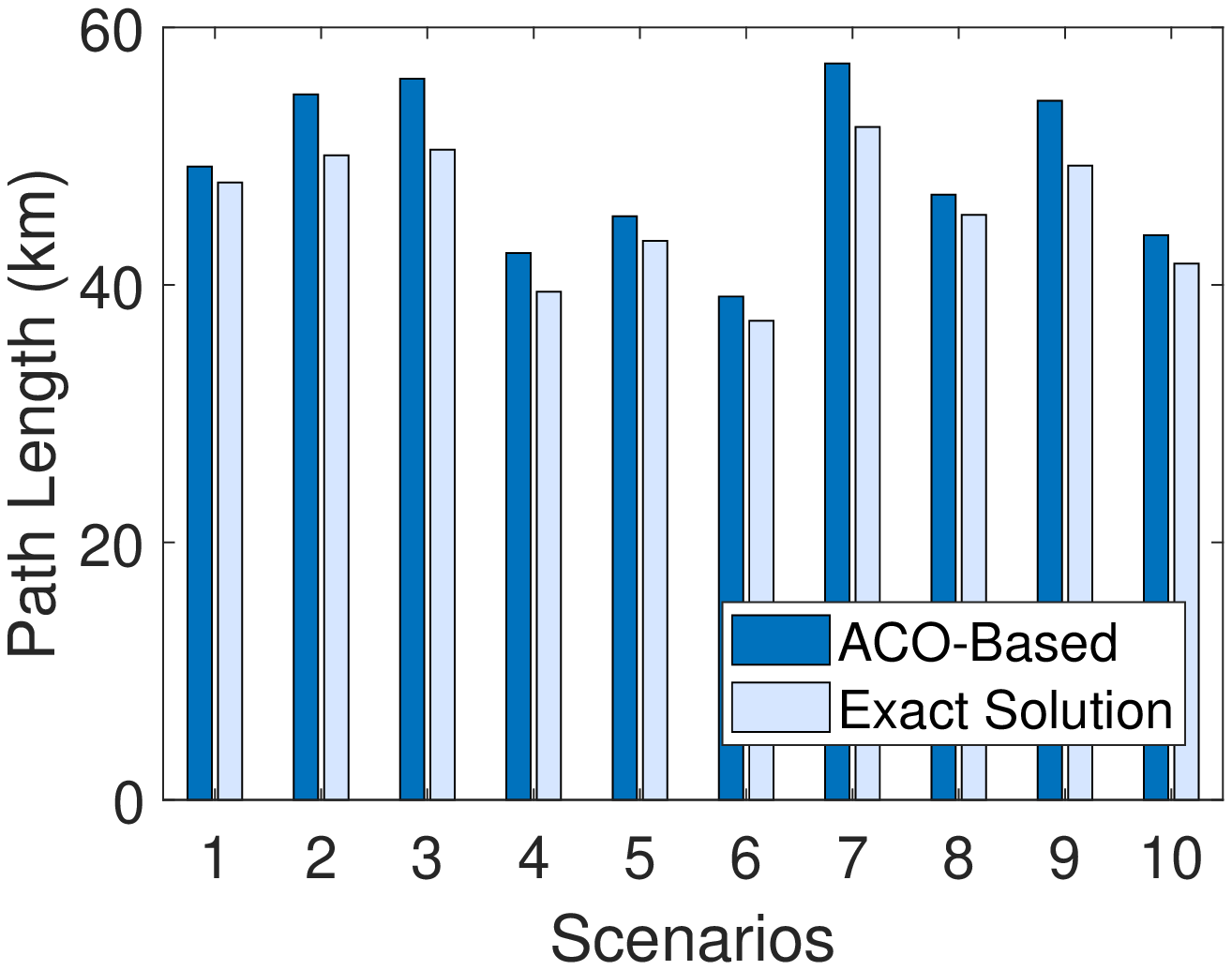}
\caption {Path length for 1 UAV.}
\label{fig:vs_ground_truth_path_len}
\end{minipage}
\hspace{1mm}
\begin{minipage}[b]{0.485\columnwidth}
\centering
\includegraphics[width=\linewidth]{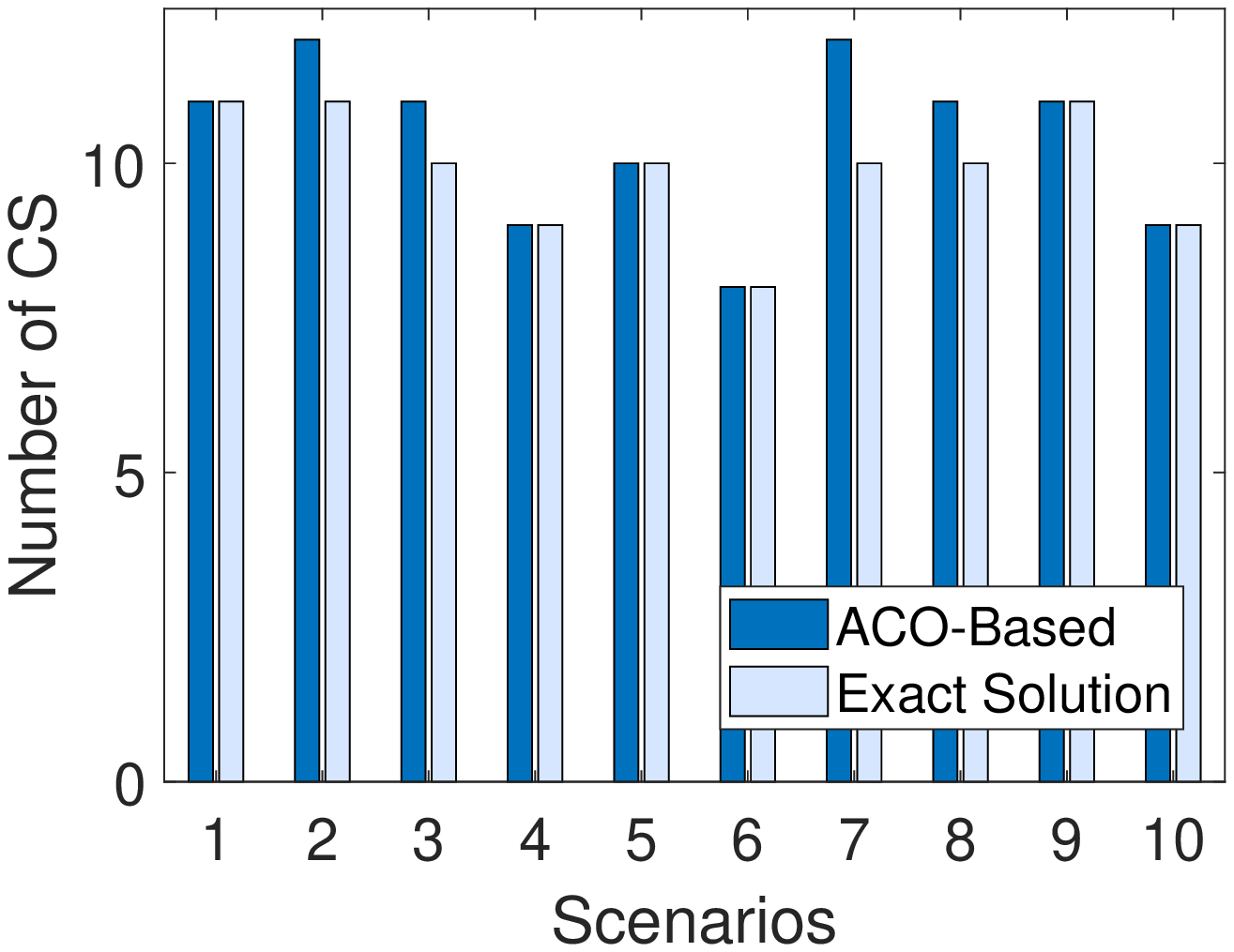}
\caption {Number of CS for 1 UAV.}
\label{fig:vs_ground_truth_num_cs}
\end{minipage}
\end{figure}

\subsection{Comparison with Exact Algorithm}
\label{sec:comparison_with_exact_solution}

The performance of UBAT was compared with the exact algorithm which finds the solution using the brute force search. A challenge for running the exact algorithm is the huge computation time. Fortunately, the true optimal solutions for a single-UAV scenario can be easily and quickly calculated~\cite{liao2016electric}. Specifically, the optimal path of a single UAV is simply the output of TSP with a set of ROIs as input, reducing the search space from 400 (number of cells) to 10 (number of ROIs). Calculating path lengths for all possible permutations for 10 ROIs completes in a few seconds. Given the shortest path, the minimum number of CS can be computed in linear time by deploying a CS at a  point on the shortest path that is farthest from the current position of a fully charged UAV~\cite{liao2016electric}.


Figures~\ref{fig:vs_ground_truth_path_len} and~\ref{fig:vs_ground_truth_num_cs} show the path length and number of CS for UBAT and the exact algorithm. On average, the path length of UBAT was 6.3\% greater with STDEV of 2.6\% compared with the true optimal solutions, and the number of CS was 4.3\% greater with STDEV of 6.0\%. Putting in different terms, UBAT had a 3.1km longer path and required 0.5 additional charging stations on average in comparison with the true optimal solutions under single-UAV scenarios.


We then measured the path length and the number of CS for multi-UAV scenarios. Unlike single-UAV scenarios, however, we cannot simply apply the TSP algorithm to find the true optimal solution for multi-UAV scenarios resulting in huge computation time. A quick, yet effective solution to this challenge is to test the algorithms with `semi' random scenarios. Specifically, 5 ROIs are randomly selected from the top of the field, and the other 5 ROIs are randomly selected from the bottom of the field such that these two sets of ROIs are individually covered by each UAV. This way the 2-UAV scenarios can be divided into two separate single-UAV scenarios that allows us to apply the TSP algorithm to obtain the true optimal solutions quickly, while UBAT is not aware of this configuration and running normally. A similar simulation setting was used for 4-UAV scenarios, \emph{i.e.,} we deployed 10 ROIs randomly in NE, NW, SE, SW regions of the field such that ROIs in each region can be independently covered by each UAV. We leave the work of obtaining the true optimal solutions for more complicated scenarios with a large number of UAVs using a high performance computing system as our future work.


\begin{figure}[!htbp]
\begin{minipage}[b]{0.485\columnwidth}
\centering
  \includegraphics[width=\linewidth]{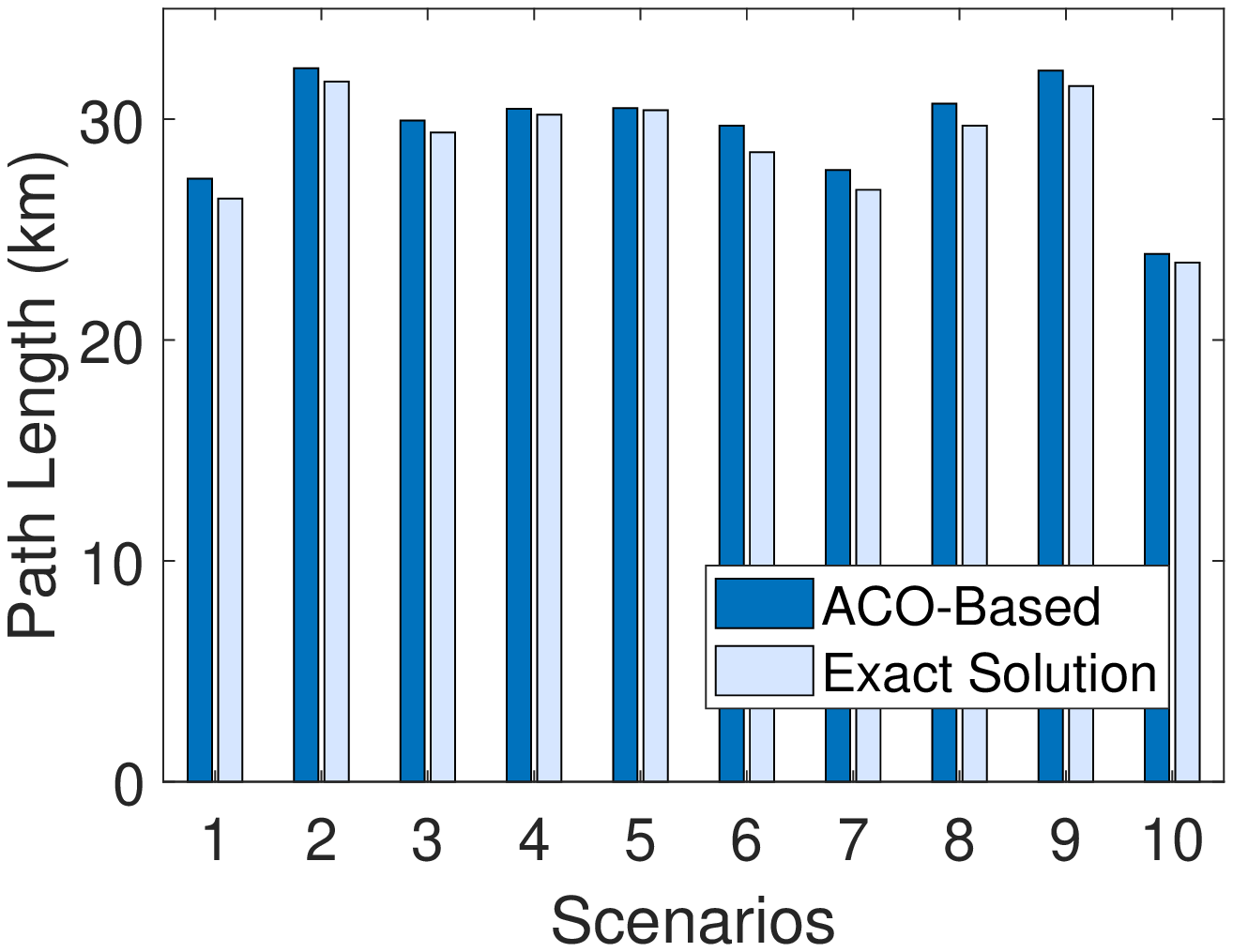}
\caption {Path length for 2 UAVs.}
\label{fig:vs_ground_truth_path_len_2UAV}
\end{minipage}
\hspace{1mm}
\begin{minipage}[b]{0.485\columnwidth}
\centering
\includegraphics[width=\linewidth]{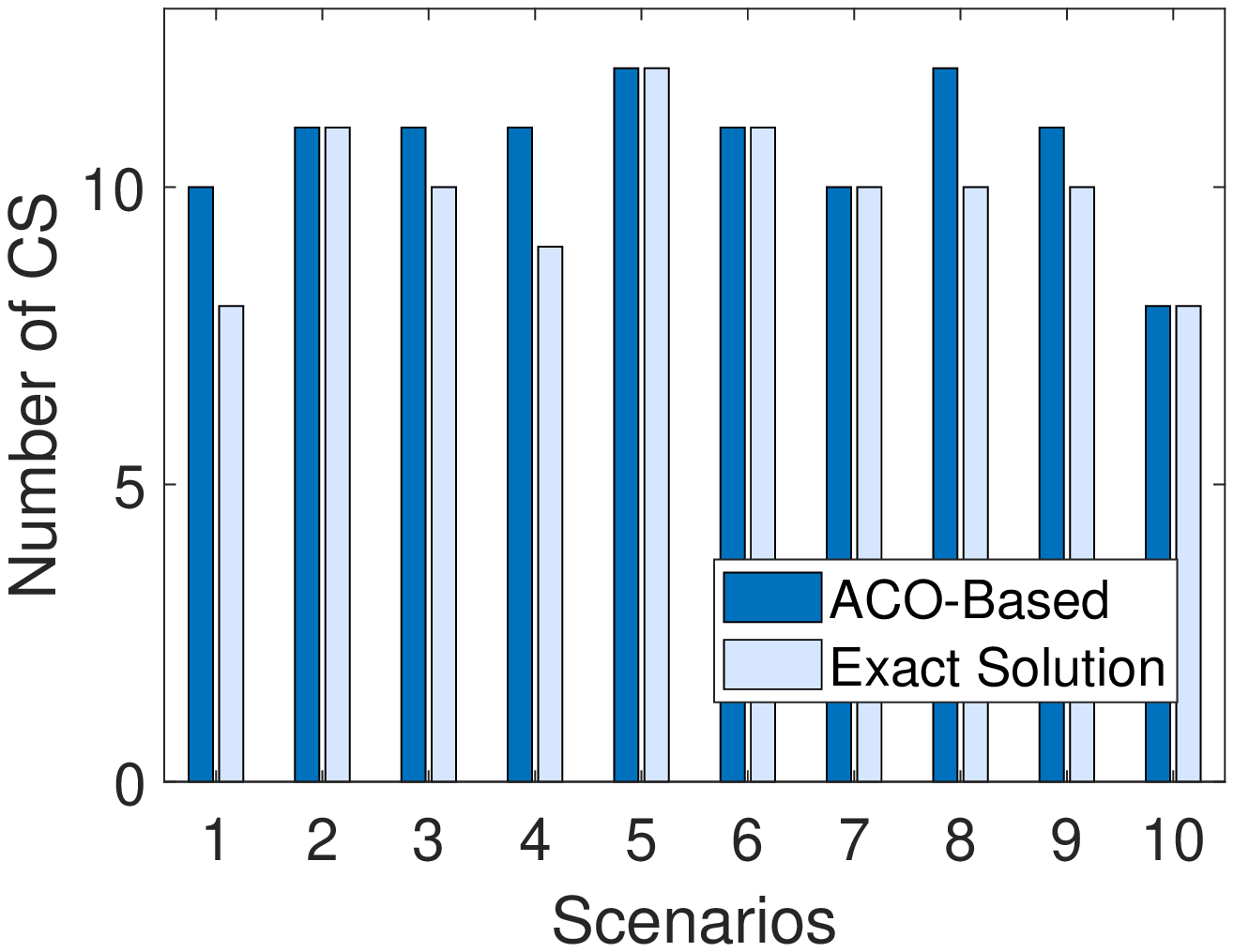}
\caption {Number of CS for 2 UAVs.}
\label{fig:vs_ground_truth_num_cs_2UAV}
\end{minipage}
\end{figure}

Figures~\ref{fig:vs_ground_truth_path_len_2UAV} and~\ref{fig:vs_ground_truth_num_cs_2UAV} show the impressive performance of UBAT for 2-UAV scenarios. The average path length of UBAT was 2.2\% greater with STDEV of 1.1\% in comparison with the exact algorithm. The average number of CS was 7.3\% greater with STDEV of 8.4\% compared with the exact algorithm. The larger percentage difference for the number of CS compared with the results for the single-UAV scenarios can be attributed to the fact that fewer charging stations were deployed for the 2-UAV scenarios because only a half of the field was covered by a UAV. Also, even though the percentage difference for the number of CS may seem much greater than that for the path length, it means only 0.8 additional charging stations on average in comparison with the true optimal solutions.

\begin{figure}[!htbp]
\begin{minipage}[b]{0.485\columnwidth}
\centering
  \includegraphics[width=\linewidth]{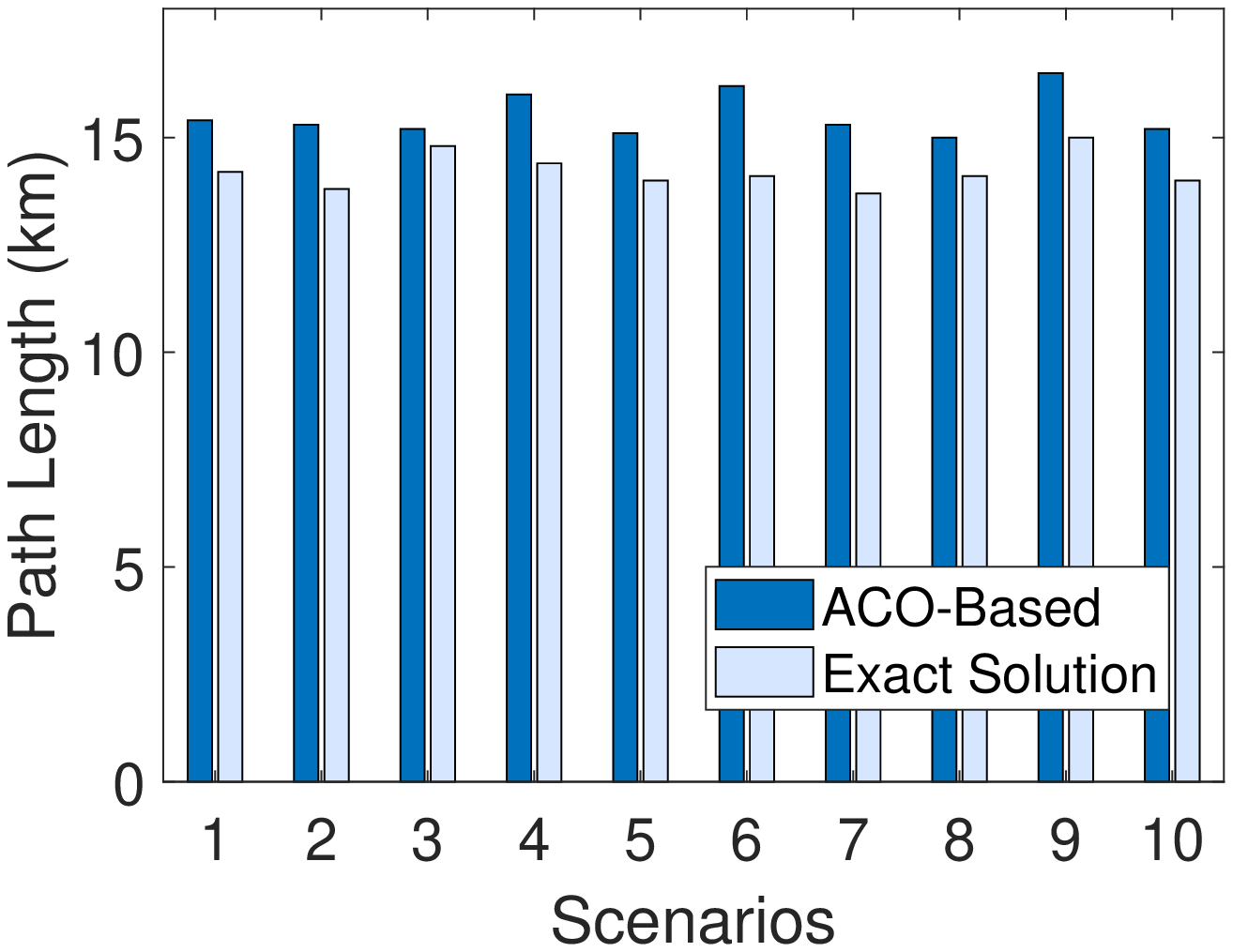}
\caption {Path length for 4 UAVs.}
\label{fig:vs_ground_truth_path_len_4UAV}
\end{minipage}
\hspace{1mm}
\begin{minipage}[b]{0.485\columnwidth}
\centering
\includegraphics[width=\linewidth]{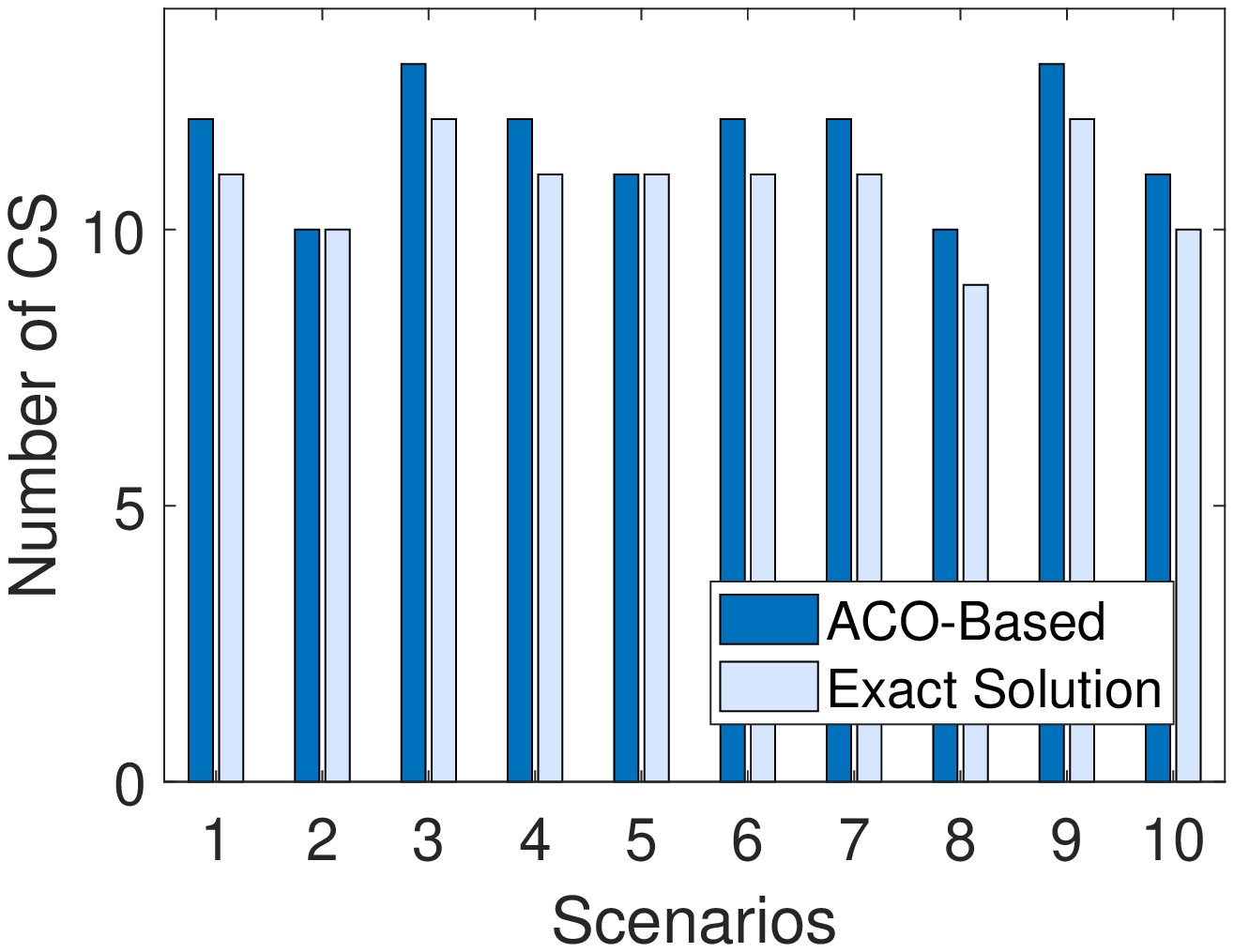}
\caption {Number of CS for 4 UAVs.}
\label{fig:vs_ground_truth_num_cs_4UAV}
\end{minipage}
\end{figure}

Similarly, Figures~\ref{fig:vs_ground_truth_path_len_4UAV} and~\ref{fig:vs_ground_truth_num_cs_4UAV} show the results for the 4-UAV scenarios. As shown, the average path length and the number of CS for the proposed framework were 8.3\% (STDEV 2.8\%) and 6.7\% (STDEV of 3.6\%) greater in comparison with the true optimal solutions, respectively. In other words, the average path length of UBAT was 1.3km longer with 0.8 additional charging stations compared with the true optimal solutions. Overall, the simulation results demonstrate that UBAT produces very high-quality solutions with slightly longer path length and less than 1 more charging stations compared with the true optimal solutions in our simulation environments.

\subsection{Effect of Enhancement Techniques}
\label{sec:effect_of_enhancement}

\begin{figure}[!htbp]
\begin{minipage}[b]{0.485\columnwidth}
\centering
\includegraphics[width=\columnwidth]{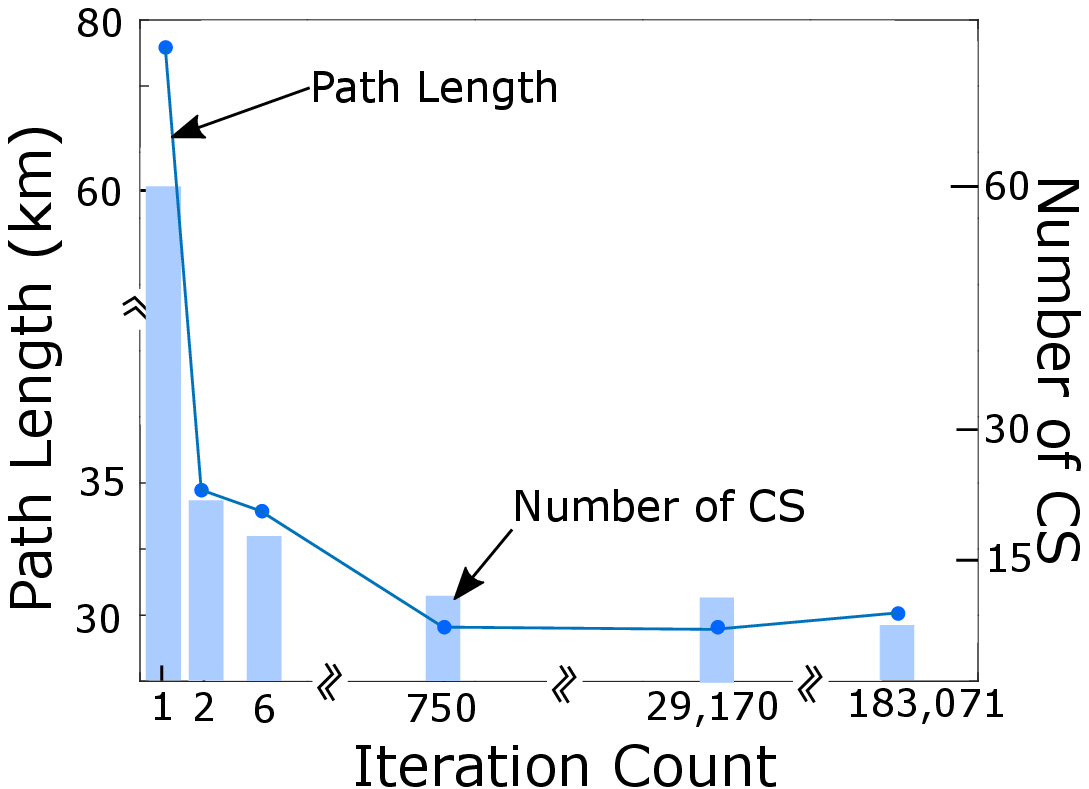}
\caption {Effect of iteration count.}
\label{fig:after_max_iteration}
\end{minipage}
\hspace{1mm}
\begin{minipage}[b]{0.485\columnwidth}
\centering
\includegraphics[width=\columnwidth]{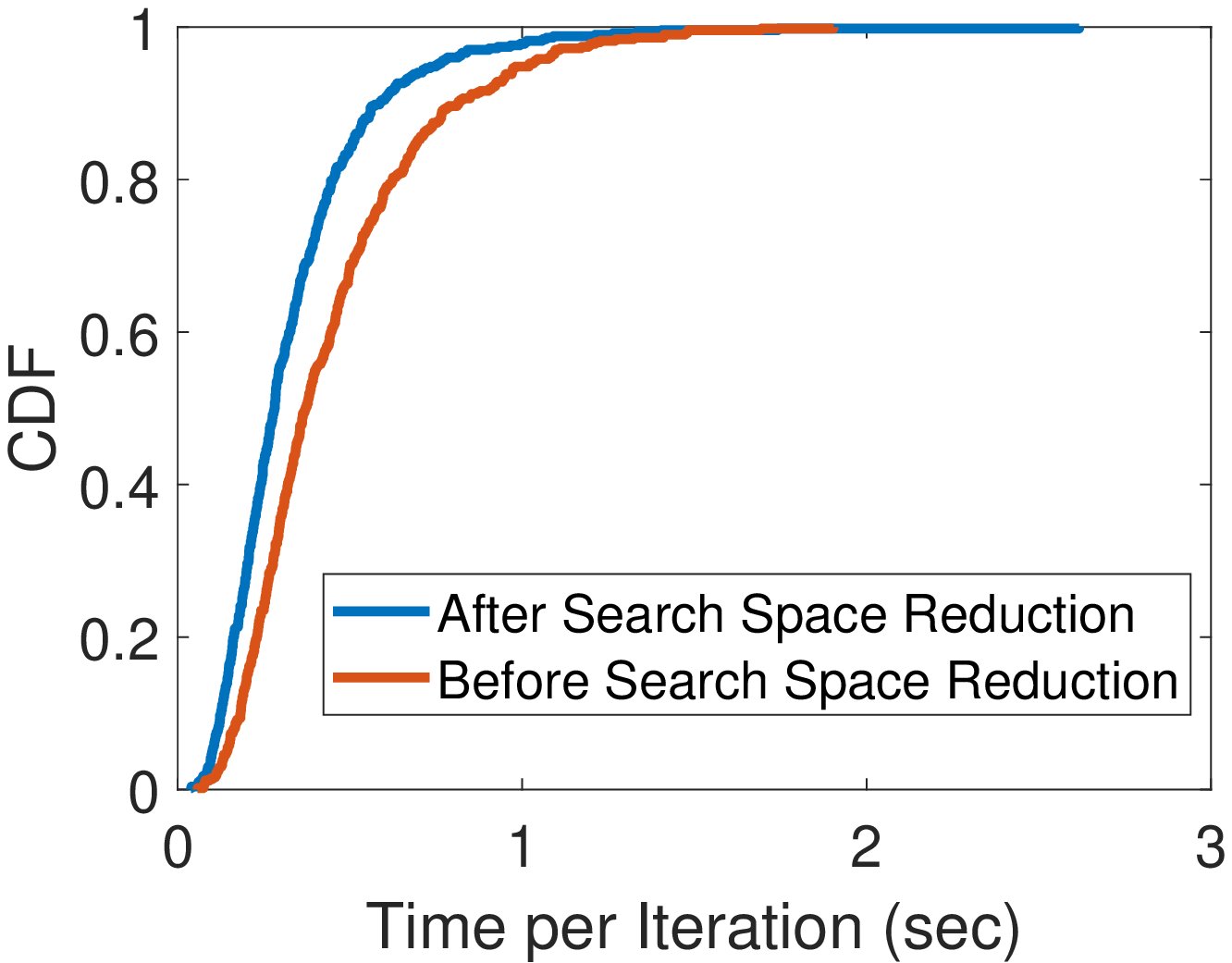}
\caption {CDF of execution time.}
\label{fig:cand_cell_opt_execution_time}
\end{minipage}
\end{figure}


\textbf{Effect of Max Iterations:} We investigate the effect of the maximum iteration count (the default is set to 30,000). It is an interesting analysis because the longer we run UBAT, the solution keeps improving. Specifically, we increased the max iteration count to 400,000 while fixing other settings for a 2-UAV scenario. We observed the following from the results (Figure~\ref{fig:after_max_iteration}): (1) a sufficiently good solution is obtained within the first few iterations; (2) after that, the solution keeps improving but very slightly. In this particular example, the path length was significantly decreased in the first 750 iterations, and it was reduced only by 0.9\% between 750th and 68,432th iterations (and virtually no improvement after that); similarly the number of CS stayed the same after the first few iterations, although in some cases, we observed that the number of CS was decreased after a very large number of iterations at the cost of the increased path length. Overall, while the solution keeps improving as we increase the maximum iteration count, the advantage is not obvious considering that major progress is made in the first few iterations and using a large maximum iteration count significantly increases the execution time.



\textbf{Effect of Search Space Reduction:} To understand the effect of the search space reduction algorithm, we measured the execution time (\emph{i.e.,} the time it takes to run a single iteration for UBAT) with and without applying the algorithm under the 2-UAV scenarios. The cumulative distribution function (CDF) of the execution time is shown in Figure~\ref{fig:cand_cell_opt_execution_time}. The results show that the average time for running a single iteration without the algorithm was 0.44 seconds, and it was decreased to 0.33 seconds when the search space reduction algorithm was used, indicating 25\% faster execution time.


\begin{figure}[!htbp]
\begin{minipage}[b]{0.485\columnwidth}
\centering
\includegraphics[width=\columnwidth]{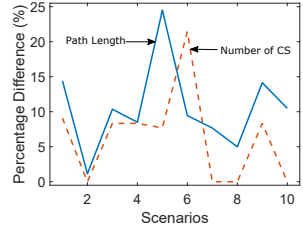}
\caption {Effect of parameter tuning.}
\label{fig:effect_of_parameter}
\end{minipage}
\hspace{1mm}
\begin{minipage}[b]{0.485\columnwidth}
\centering
\includegraphics[width=\columnwidth]{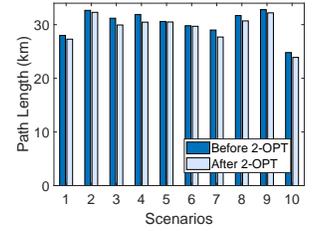}
\caption {Effect of path improvement.}
\label{fig:effect_of_2opt}
\end{minipage}
\end{figure}


\textbf{Effect of Parameter Tuning:} We evaluate the performance of the parameter tuning algorithm. We started with some arbitrary values for the parameters $Q_1$ and $Q_2$. We then executed UBAT only for a few iterations, \emph{i.e.}, 1,000 iterations. As observed in the experiments for the max iteration count, a sufficiently good solution is obtained in the first few iterations. The obtained intermediate solution, \emph{i.e.,} the path length, and the number of CS were used as new values for parameters $Q_1$, and $Q_2$, respectively. We then ran UBAT with the new parameters and compared the results with those obtained with arbitrarily selected parameters. The results in Figure~\ref{fig:effect_of_parameter} show that the path length and the number of CS were improved by up to 24.5\% and 21.4\% (10.5\% and 6.3\% on average), respectively, when the parameter tuning algorithm was applied.

We then evaluate the performance gain when the parameter tuning algorithm is applied repeatedly. An interesting observation was that the performance gain was very marginal even from the second time running the parameter tuning algorithm. Specifically, the path length and the number of CS were improved only by up to 0.3\% on average. In the third time for tuning the parameters, virtually no improvement was observed, indicating that running the parameter tuning process once or twice suffices to obtain good parameters. Note that all simulation results were obtained using this parameter tuning method, showing the effectiveness and practicality of the proposed approach.




\textbf{Effect of Path Correction:} We analyze the effect of the path improvement algorithm. We measured the path length with and without applying the algorithm, while fixing all other settings for the 2-UAV scenarios. It should be noted that the number of CS was not measured because the path improvement algorithm does not increase the number of CS at the cost of improving the path. The results are depicted in Figure~\ref{fig:effect_of_2opt}. The path length was decreased by 2.5\% on average with STDEV of 1.6\% when the path improvement algorithm was applied. The results indicate that by skipping low-quality solutions (\emph{e.g.,} paths with self-crossings), the search space is explored more effectively, leading to better solutions.

\section{Conclusion}
\label{sec:conclusion}

We have presented UBAT, an optimization framework that jointly optimizes UAV trajectories and locations of charging stations. Enhancement techniques are developed to improve the convergence time and quality of solutions. Through extensive simulations, we demonstrate that UBAT effectively calculates the UAV trajectories and placement of charging stations. The future work is to perform theoretical analysis on the algorithm complexity regarding the number of UAVs, field size, and scale of discrete grids and to conduct experiments based on actual UAV robotic platforms.


\bibliographystyle{IEEEtran}
\bibliography{mybibfile}



\begin{IEEEbiography}[{\includegraphics[width=1in,height=1.25in,clip,keepaspectratio]{won}}]{Myounggyu Won}
(M'13) received a Ph.D. degree in Computer Science from Texas A\&M University at College Station, in 2013. He is an Assistant Professor in the Department of Computer Science at the University of Memphis, Memphis, TN, United States. Prior to joining the University of Memphis, he was an Assistant Professor in the Department of Electrical Engineering and Computer Science at the South Dakota State University, Brookings, SD, United States from Aug. 2015 to Aug. 2018, and he was a postdoctoral researcher in the Department of Information and Communication Engineering at Daegu Gyeongbuk Institute of Science and Technology (DGIST), South Korea from July 2013 to July 2014.  His research interests include smart sensor systems, connected vehicles, mobile computing, wireless sensor networks, and intelligent transportation systems. He received the Graduate Research Excellence Award from the Department of Computer Science and Engineering at Texas A\&M University - College Station in 2012.
\end{IEEEbiography}

\end{document}